\documentclass[copyright,creativecommons]{eptcs}
\providecommand{\event}{FM MDD 2017}

\usepackage{underscore}           
\usepackage{amsmath}
\usepackage{amssymb}
\usepackage{graphicx}
\usepackage {bsymb,b2latex,mybcode}

\begingroup
\catcode`_=13
\gdef\var{%
	\begingroup
	\catcode`_=13
	\def_{\!\_}%
	\dovar
}
\gdef\dovar#1{%
	\mathit{#1}%
	\endgroup
}
\endgroup

\begin{document}
	
\title{Towards Integrated Modelling of \\ Dynamic Access Control with UML and Event-B}
\author{Inna Vistbakka \qquad\qquad Elena Troubitsyna
\institute{\AA{}bo Akademi University,\\ Turku, Finland\\}
\email{inna.vistbakka@abo.fi \qquad\qquad elena.troubitsyna@abo.fi}
}
\def\titlerunning{Towards Integrated Modelling of Dynamic Access Control with UML and Event-B}
\def\authorrunning{Inna Vistbakka and Elena Troubitsyna}

\maketitle

\begin{abstract}
Role-Based Access Control (RBAC) is a popular authorization model used to manage data-access constraints in a wide range of systems. RBAC usually defines the static view on the access rights. However, to ensure dependability of a system, it is often necessary to model and verify state-dependent access rights. Such a modelling allows us to explicitly define the dependencies between the system states and permissions to access and modify certain data. In this paper, we present a work-in-progress on combining graphical and formal modelling to specify and verify dynamic access control. The approach is illustrated by a case study -- a reporting management system.

\end{abstract}

\section{Introduction}
\label{sec:intro}
The Role-Based Access Control (RBAC) is a de-facto standard mechanism for specifying the data access policies in a large variety of computer-based systems \cite{FerraioloSGKC01}.  A role represents a job function in the context of an organization and has associated authorities. The RBAC policy of a system specifies the authorities granted to each role. Usually RBAC gives a static view on the access rights associated with each role, i.e., it defines the permissions to manipulate certain data without referring to the system state. However, such a static view is often insufficient for ensuring system dependability because it might undermine dynamic data integrity properties. In this paper, we present a work-in-progress aiming at combining graphical and formal modelling to represent the dynamic access policy.

A graphical representation is often used as a front-end of a formal model \cite{SereT99,SereT99FM,Troubitsyna03IPDPS,Troubitsyna08}. It facilitates a transition from an informal, natural language, requirements description to their formal representation. In our approach, graphical modelling is used to represent the system functions and business logic associated with them. We rely on the use case model to represent the system roles and functions. The activity diagram shows the workflow  associated with the defined functions. From the graphical representation we make a transition to a formal model. We use Event-B  \cite{EventB} as our formal modelling framework. Event-B is a state-based modelling notation. It supports a top-down development approach to correct-by-construction system development. The system development consists of a sequence of correctness-preserving model transformations -- refinements. Correctness of models and refinements can be verified by proofs. Rodin platform  \cite{RODINPLAT} automates modelling and verification in Event-B.

While creating Event-B specifications, we consult the graphical models to define the roles and the corresponding functions. The dynamic access rights are modelled as enabling or disabling certain operations on data, which depend on the system state. Formal modelling allows us to ensure  that the dynamic access rights are preserved throughout the system workflow.

The paper is structured as follows: in Section 2, we define the main concepts of dynamic access policy and present our case study -- reporting management system.  We define the corresponding graphical models modelling roles, use cases and the workflow. In Section 3, we give a brief introduction into Event-B. In Section 4, we present the formal models that have been constructed on the basis of the graphical models. We discuss how to represent the dynamic access rights in Event-B. Finally, in Section 5, we overview the related work and discuss the proposed approach.

\section{Towards Reasoning About Dynamic Access Policy }
\label{sec:reasoning}
In this section, we will define the main notions and primitives required to reason about behaviour of a system using the dynamic access control model.  We will discuss the notions of users and their roles, data and the rights to manipulate data. We will also introduce some functions and relations to define the required inter-relationships between these notions.

Let $USERS=\{u_1, u_2, ..., u_n\}$ be a set of users.  The users are core elements of the system. In general, a user may stand for a person in the organisation, an administrative entity or a non-person entity, such as a computing (sub)system. We use the term user to cover all the cases.

Let $ROLES=\{r_1, r_2, ..., r_k\}$ be a set of possible user roles within the system. A role is usually seen as a job function performed by a user within an organisation. For example, in a security model, a role is used to indicate the job-related access rights to the data.

Let $RIGHTS=\{ri_1, ri_2, ..., ri_m\}$ be a set of basic access rights that can be defined over the data, e.g., \textit{create, read, write} rights. Moreover,  $DATA=\{d_1, d_2, ..., d_l\}$ denotes a set of data entities.
The users can access data only by executing operations on a data entity that are regulated by corresponding basic rights. Operations and data are predefined by the underlying system for which RBAC is specified.

The users can access data based on the set of assigned roles.
A user authorisation list can be defined as the mapping between users and roles:
$$\var{UR_Rel} : USERS \rightarrow {\pow}(ROLES),$$ which assigns a set of user roles to a  given user.
The notation ${\pow}(ROLES)$ stands for the powerset (set of all subsets) type over elements of the type  $ROLES$.

To formally define access rights to the data provided by the system to different user roles, we define a function $\var{RR_Rel}$ that maps each user  role to a set of the allowed rights:
$$\var{RR_Rel} : ROLES \rightarrow {\pow}(RIGHTS).$$

The above definition describes what a user with a specific role is allowed to do with data in general. However, it does not take into account the system workflow, i.e., it abstracts away from the fact that the access rights over a data entity for each role also depend on the system state. To demonstrate how this issue can be addressed via formal modelling and verification in Event-B, let us now consider an example of a system with a dynamic access control. Then we will show how we can create a specification of such a system and prove the correctness of its behaviour.


\paragraph{A Case Study -- Reporting Management System.}
\label{sec:cs}
To illustrate modelling opportunities of Event-B with combination of UML, we use a simple case study -- a reporting management system.

The Reporting Management System (RMS) is used by different kinds of employees in an organisation to send periodic (e.g., monthly) work reports.
The system has the following requirements:
\begin{itemize}
 \item \textit{The system has three roles}  -- reporter, controller and  administrator;
 \begin{itemize}
 	\item Every reporter is associated with (supervised by) some controller;
 	\item Every controller can have more than one associated  reporter;
 	\item Every controller is associated with (supervised by) one administrator;
    \item Every  administrator can have more than one associated controller.
 	 \end{itemize}
 \vspace*{0.2cm}
 \item \textit{Functionality associated with the roles}:
 \begin{itemize}
 \item A reporter can create a new report, modify an existing report or delete a not-approved report, as well as submit a report to its associated controller;
 \item A controller can read submitted report received from one of the associated reporters, and  can either approve or disapprove it;
 \item An administrator has an access to all the reports of all her/his associated controllers, and it is her/his responsibility to register reports approved by the controllers.
 \end{itemize}
\vspace*{0.2cm}
 \item \textit{Report access policies}:
 \begin{itemize}
 \item Until a report is submitted, the reporter can modify or delete it.
 \item As soon as a report is submitted, it cannot be altered or deleted by the reporter any longer.
 \item Upon controller's approval, the report is registered by the administrator.
 \item In case of disapproval, the report is returned back to the reporter and can be further modified or deleted.
 \end{itemize}
\end{itemize}

\begin{figure}[b!]
	\centering
	\includegraphics[width=0.7\textwidth]{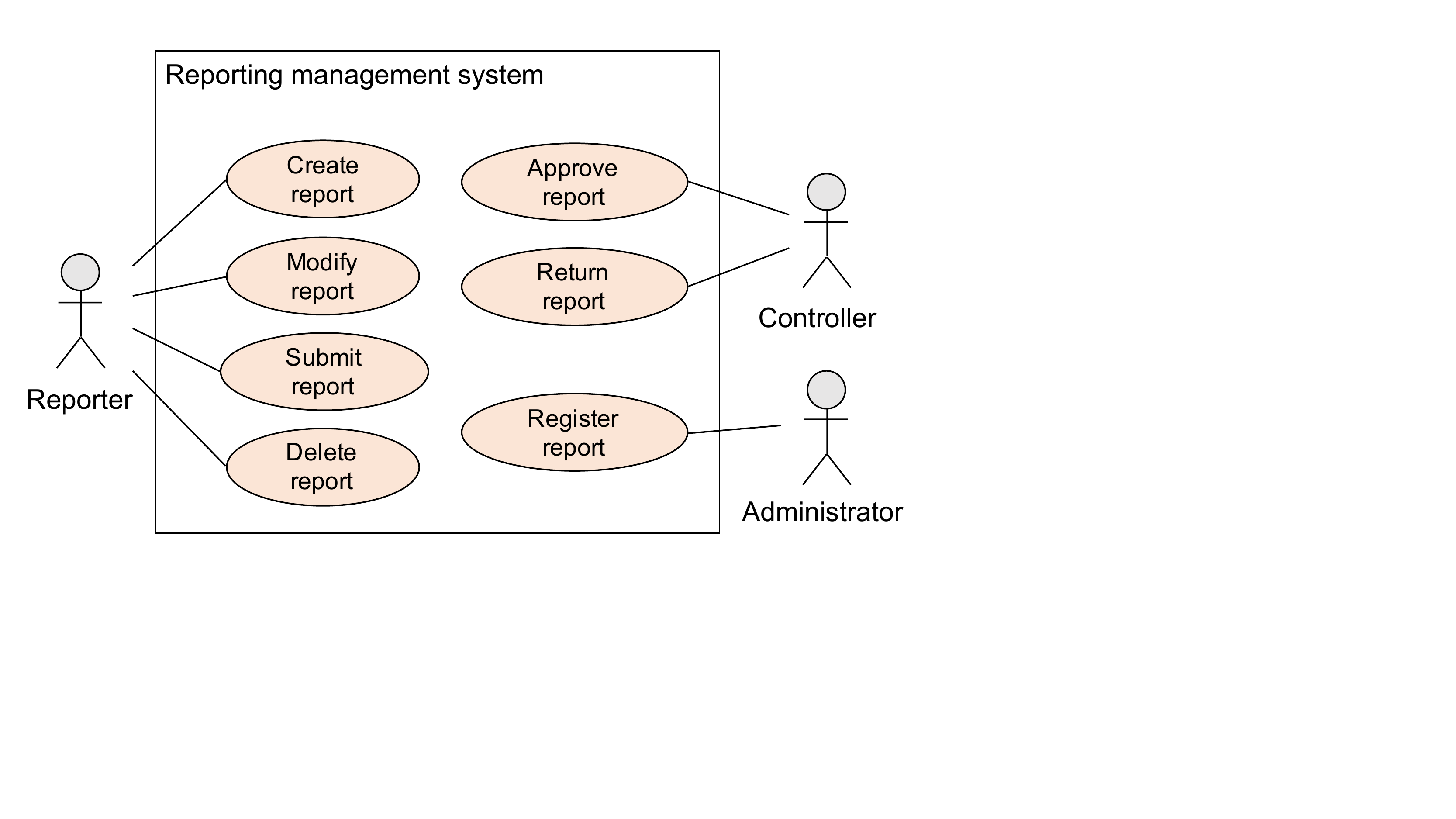}
	\caption{Use Case Diagram -- Reporting Management System}
	\label{fig:UseCaseReportronic}
\end{figure}

\begin{figure}[h]
	\centering
	\includegraphics[width=0.55\textwidth]{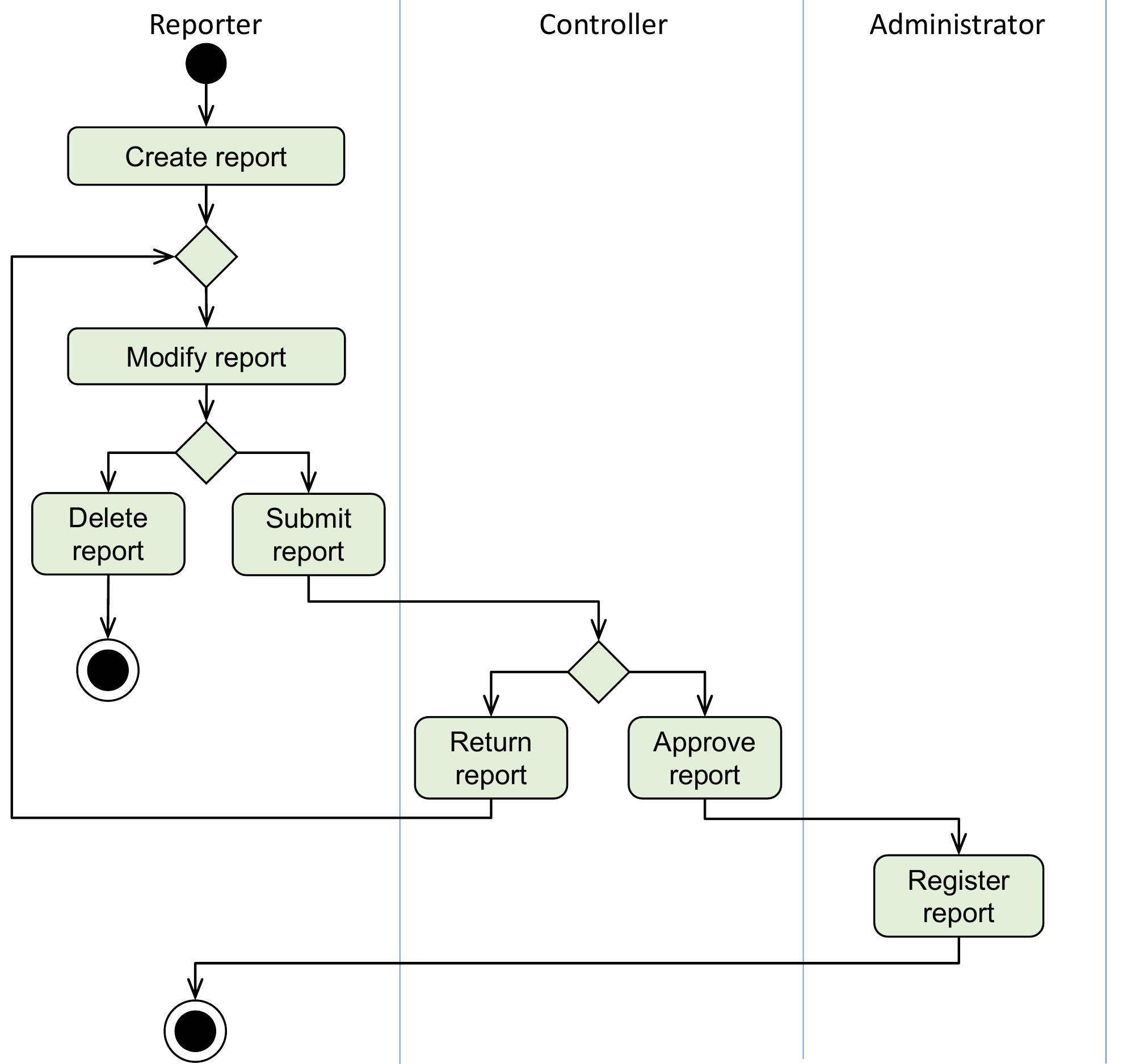}
	\caption{Activity Diagram -- Reporting Management System }
	\label{fig:ActivityDiagramReportronic}
\end{figure}

The use case diagram of RMS  is presented in Figure~\ref{fig:UseCaseReportronic}. It shows the actors, their roles in the system and also their possible interactions with the system. The  activity diagram for RMS's workflow  is presented in Figure~\ref{fig:ActivityDiagramReportronic}.

Let us note that each actor function requires certain basic access rights. For instance, a reporter, to be able to execute \textit{Modify} function, should have \textit{read} and  \textit{write}  access rights to the report file.  In its turn, a reporter's supervisor --  controller  --   should have  \textit{read} and \textit{write} access rights to the same file to execute \textit{Approval} operation. However, as soon as a reporter submit a report to a controller, she/he can have only  \textit{read} access right to the report file.

In Section~\ref{sec:dev}, we will present a refinement-based development of RMS. We will discuss how to represent dynamic access rights in Event-B, formally specify  all actors' operations over the data and prove the correctness of behaviour of RMS.
Before that, we will briefly overview Event-B modelling framework and its refinement approach.

\section{Modelling and Refinement in Event-B}
\label{sec:eventb}

Event-B \cite{EventB} is a state-based framework that promotes the correct-by-construc\-tion approach to system development and formal verification by theorem proving. In Event-B, a system model is specified using the notion of an \emph{abstract state machine} \cite{EventB}. An abstract state machine encapsulates the model state, represented as a collection of variables, and defines operations on the state, i.e., it describes the dynamic behaviour of a modelled system.
A machine also has an accompanying component, called \emph{context}, which includes user-defined sets, constants and their properties given as model axioms.

\begin{figure}[h]
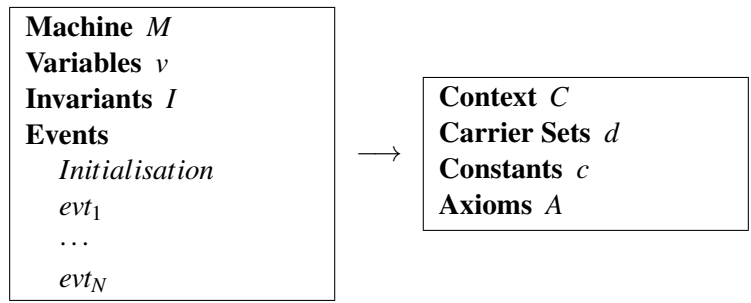

\begin{center}
\fbox{
\parbox[center]{4cm}{
\textbf{Machine} $\;M$ \\
\hspace*{0pt}\textbf{Variables} $\; v$ \\
\hspace*{0pt}\textbf{Invariants} $\; I$ \\
\hspace*{0pt}\textbf{Events} \\
\hspace*{10pt} $Initialisation$ \\
\hspace*{10pt} $evt_1$ \\
\hspace*{10pt} $\cdots$ \\
\hspace*{10pt} $evt_N$
}}
\hskip 5pt
$\longrightarrow$
\hskip 5pt
\fbox{
\parbox[center]{4cm}{
\textbf{Context} $\;C$ \\
\hspace*{0pt}\textbf{Carrier Sets} $\; d$ \\
\hspace*{0pt}\textbf{Constants} $\; c$ \\
\hspace*{0pt}\textbf{Axioms} $\; A$
\smallskip
}}
\caption{Event-B machine and context}
\label{fig:am}
\end{center}
\end{figure}

A general form for Event-B models is given in Figure~\ref{fig:am}.
The machine is uniquely identified by its name $M$. The state variables, $v$, are declared in the \textbf{Variables} clause and initialised in the $Initialisation$ event. The variables  are strongly typed by the constraining predicates $I$ given in the \textbf{Invariants} clause. The invariant clause might also contain other predicates defining essential properties (e.g., safety invariants) that should be preserved during system execution.

The dynamic behaviour of the system is defined by a set of atomic \textit{events}. In general, an event has the following form:
$$
e\;\widehat{=}\;\textbf{any } a \textbf{ where } G_{e} \textbf{ then } R_{e} \textbf{ end},
$$

\noindent where $e$ is the event's name, $a$ is the list of local variables, the \textit{guard} $G_{e}$ is a predicate over the local variables of the event and the state variables of the system.  The body of an event is defined by a \textit{multiple} (possibly nondeterministic) assignment over the system variables. In Event-B, an assignment represents a corresponding next-state relation $R_{e}$. Later on, using the concrete syntax in our Event-B models, we will rely on two kinds of assignment statements: deterministic ones, expressed in the standard form $x := E(x,y)$, and non-deterministic ones, represented as $x :\mid \var{some_condition}(x,y,x')$. In the latter case, the state variable $x$ gets non-deterministically updated by the value $x'$, which may depend on the initial values of the variables $x$ and $y$.

 The guard defines the conditions under which the event is \emph{enabled}, i.e., its body can be executed.  If several events are enabled at the same time, any of them can be chosen for execution nondeterministically.

 If an event does not have local variables, it can be described simply as:
 $$
 e\;\widehat{=}\;\textbf{when } G_{e} \textbf{ then } R_{e} \textbf{ end}.
 $$

Event-B employs a top-down refinement-based approach to system development. Development typically starts from an abstract specification that nondeterministically models the most essential functional requirements. In a sequence of refinement steps, we gradually reduce nondeterminism and introduce detailed design decisions. In particular, we can add new events, split events as well as replace abstract variables by their concrete counterparts, i.e., perform \emph{data refinement}.

The consistency of Event-B models, i.e., verification of well-formedness and invariant preservation as well as correctness of refinement steps, is demonstrated by discharging a number of verification conditions -- proof obligations.
For instance, to verify \textit{invariant preservation}, we should prove the following logical formula:
\begin{equation*}
A(d,c),\ I(d,c,v),\ G_e(d,c,x,v),\ R_e(d,c,x,v,v')\ \vdash\ I(d,c,v'),
\tag*{\text{(INV)}}
\end{equation*}
where $A$ are the model axioms, $I$ are the model invariants, $d$ and $c$ are the model constants and sets respectively, $x$ are the event's local variables and $v, v'$ are the variable values before and after event execution. The full definitions of all the proof obligations are given in \cite{EventB}.

The Rodin platform \cite{RODINPLAT} provides an automated support for formal modelling and verification in Event-B. In particular, it automatically generates the required proof obligations and attempts to discharge them. The remaining unproven conditions can be dealt with by using the provided interactive provers.

\section{Formal Development of the Reporting Management System in Event-B}
\label{sec:dev}
In this section, we will present a refinement-based development of the reporting management system discussed in Section~\ref{sec:reasoning}. We will model all actors' functions as Event-B events. In general, the semantics of each Event-B event is unambiguously defined by a binary relation between all possible \textit{pre-} and \textit{post-states} of the event. Therefore, modelling RMS in Event-B would allow us to define pre- and post-states for every operation with the data, and model the overall system behaviour as a state transition system.

Moreover, our development will incorporate the essential concepts and relationships between system elements described in Section~\ref{sec:reasoning}.  We will separately discuss both static and dynamic system aspects, represented by Event-B contexts and machines respectively.

\begin{figure}[h!]
	\centering
	\includegraphics[width=0.7\textwidth]{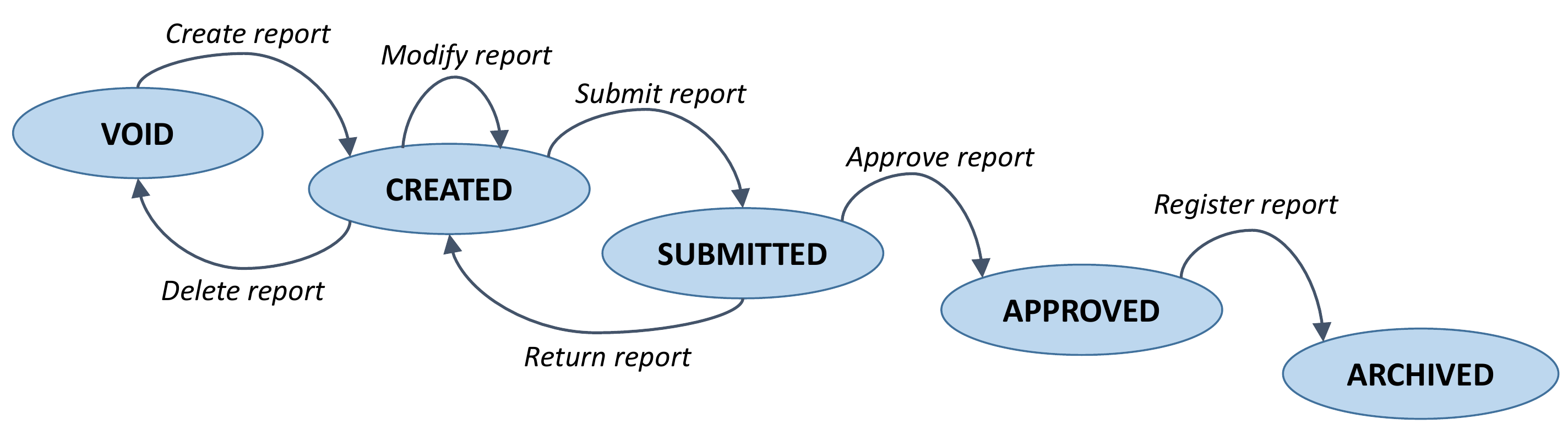}
	\caption{State diagram -- Reporting Management System}
	\label{fig:StateDiagram}
\end{figure}

\paragraph{\textbf{Reporting Management  System: Abstract Model.}}
Let us note that each actor's function changes the state of a certain report. Hence, the overall behaviour of the system, for each particular report,  can be considered as a set of transitions between all the possible states of the report. The corresponding state diagram is represented in Figure~\ref{fig:StateDiagram}.

In our initial Event-B specification, we abstractly specify changing of states of reports in our system. We model the set of reports in the system as a function $\var{report_state}$. Initially each report has the state $VOID$. The actual report creation is modelled by the event $\mathsf{CreateReport}$ that changes the state of a single report $rp$ to $CREATED$, i.e.,  $\var{report_state}(rp) := CREATED$. Then the events $\mathsf{ModifyReport}$, $\mathsf{DeleteReport}$ and $\mathsf{SubmitReport}$ become enabled. When the report is submitted, its state changes to $SUBMITTED$. Upon report approval, its state is changed to $APPROVED$, otherwise, if the report is rejected, it returns back to the state  $CREATED$.  Finally, once the administrator registers already approved report, the report goes to its final state $ARCHIVED$.

\begin{figure}[t!]
\centering
\begin{center}
\begin{small}
\fbox{
\parbox[center]{13cm}{
\hspace*{0pt} $\textbf{Machine}$ $\mathsf{RMS\_m0}$ \vspace{1pt}\\
\hspace*{0pt} $\textbf{Sees}$ $\mathsf{RMS\_c0}$ \vspace{1pt}\\
\hspace*{0pt} $\textbf{Variables}  \;  \textit{reports},  \var{report_state}$ \vspace{1pt}\\
\hspace*{0pt} $\textbf{Invariants} $\\
\hspace*{47pt} $ reports \subseteq REPORTS  \; \land $\\
\hspace*{47pt} $\var{report_state}  \in  REPORTS  \tfun  \textit{REPORT\_STATES} \;  \land $   \\
\hspace*{47pt} $\var{report_state}[reports] \subseteq \textit{REPORT\_STATES} \setminus \{VOID\} ...$\\
\hspace*{0pt} $\textbf{Events}\vspace*{5pt}$   \\
\parbox[center]{6.5cm}{
\hspace*{0pt} $\mathsf{Initialisation}$ $ \;\widehat{=}$ ... \vspace*{2pt} \\
\hspace*{0pt} $\mathsf{CreateReport}  \;\widehat{=}$ \vspace{1pt} \\
\hspace*{5pt} $\textbf{any} \; \; rp$\\
\hspace*{5pt} $\textbf{where} \; \; rp  \in \textit{DATA}$\\
\hspace*{5pt} $\var{report_state}(rp)=\textit{VOID}$\\
\hspace*{5pt} $\textbf{then}$\\
\hspace*{5pt} $\textit{reports} \bcmeq{}  \textit{reports} \bunion{} \{rp\}$\\
\hspace*{5pt} $\var{report_state}(rp)   \bcmeq{} \textit{CREATED}$\\
\hspace*{5pt} $\textbf{end}\vspace{3pt}$\\
\hspace*{0pt} $\mathsf{ModifyReport}\;\widehat{=}$ \vspace{1pt} \\
\hspace*{5pt} $\textbf{any} \; \; rp$\\
\hspace*{5pt} $\textbf{where} \; \; rp \in \textit{reports}$\\
\hspace*{5pt} $\var{report_state}(rp) : = \textit{CREATED}$\\
\hspace*{5pt} $\textbf{then}$\\
\hspace*{5pt} $skip$\\
\hspace*{5pt} $\textbf{end}\vspace{3pt}$\\
\hspace*{0pt} $\mathsf{DeleteReport}\;\widehat{=}$ ...\vspace{2pt}  \\
\hspace*{5pt} $\textbf{any} \; \; rp$\\
\hspace*{5pt} $\textbf{where} \; \; rp \in \textit{reports}$\\
\hspace*{5pt} $\var{report_state}(rp) = \textit{CREATED}$\\
\hspace*{5pt} $\textbf{then}$\\
\hspace*{5pt} $\var{report_state}(rp) : = \textit{VOID}$\\
\hspace*{5pt} $\textit{reports} \bcmeq{}  \textit{reports} \setminus\{rp\}$\\
\hspace*{5pt} $\textbf{end}$\\

}
\parbox[center]{6.5cm}{
\hspace*{0pt} $\mathsf{SubmitReport}\;\widehat{=}$  \vspace{1pt} \\
\hspace*{5pt} $\textbf{any} \; \; rp$\\
\hspace*{5pt} $\textbf{where} \; \; rp \in \textit{data}$\\
\hspace*{5pt} $\var{report_state}(rp) = \textit{CREATED}$\\
\hspace*{5pt} $\textbf{then}$\\
\hspace*{5pt} $\var{report_state}(rp) := \textit{SUBMITTED}$\\
\hspace*{5pt} $\textbf{end}\vspace{2pt}$\\
\hspace*{0pt} $\mathsf{ApproveReport}\;\widehat{=}$ \vspace{1pt} \\
\hspace*{5pt} $\textbf{any} \; \; rp$\\
\hspace*{5pt} $\textbf{where} \; \; rp \in \textit{reports}$\\
\hspace*{5pt} $\var{report_state}(rp) = \textit{SUBMITTED}$\\
\hspace*{5pt} $\textbf{then}$\\
\hspace*{5pt} $\var{report_state}(rp) \bcmeq{}  \textit{APPROVED}$\\
\hspace*{5pt} $\textbf{end}\vspace{2pt}$\\
\hspace*{0pt} $\mathsf{ReturnReport}\;\widehat{=}$ ... \vspace{2pt} \\
\hspace*{0pt} $\mathsf{RegisterReport}\;\widehat{=}$\\
\hspace*{5pt} $\textbf{any} \; \; rp$\\
\hspace*{5pt} $\textbf{where} \; \; rp \in \textit{reports}$\\
\hspace*{5pt} $\var{report_state}(rp) =  \textit{APPROVED}$\\
\hspace*{5pt} $\textbf{then}$\\
\hspace*{5pt} $\var{report_state}(rp) \bcmeq{}  \textit{ARCHIVED}$\\
\hspace*{5pt} $\textbf{end}$\\
$\textbf{end}$}
}}
\end{small}
\caption{The machine $\mathsf{RMS\_abs}$}
\label{fig:devReportronicAbs}
\vspace*{-0.2cm}
\end{center}
\vspace*{-0.3cm}
\end{figure}

The structure of our initial model -- machine $\mathsf{RMS\_abs}$ -- is given in Figure~\ref{fig:devReportronicAbs}. This machine essentially covers all the use cases presented in Figure~\ref{fig:UseCaseReportronic}.  All required sets and constants are defined in the static part of the model -- context $\mathsf{RMS\_c0}$ (not presented in the paper).

\paragraph{\textbf{First Refinement: Introducing Roles.}}
The purpose of this refinement step is to elaborate on the initial system specification and introduce user roles. We will link each role with the set of functions that correspond to it. Moreover, for each role, we will define the required basic access rights -- \textit{create, read, write, delete}.

In the context part of Event-B specification, we define a set of actor roles  $\textit{ROLES}=\{Reporter, \linebreak Controller, \; Administrator\}$. $\var{RIGHTS}$ is the set of basic access rights, where  $\var{RIGHTS}= \{C, \; R, \; W, \; D\}.$

To specify dynamic access rights for the introduced  roles, we define a variable  $permissions$ with the following properties:
$$ permissions \in ROLES\times REPORTS  \tfun \pow(\var{RIGHTS}),$$
\begin{multline*}
\forall r\in REPORTS \cdot permissions(Reporter, r) \subseteq \{C, W, R, D \}\, \land \\
permissions(Controller, r) \subseteq \{R,W\} \land permissions(Administrator, r) \subseteq \{R,W\}.
\end{multline*}

\noindent The variable $permissions$ is a function that assigns to each role and a report a number of possible  access rights that can be associated with the role.

Obviously, for each role, the set of available access rights to a report depends on the current state of this report. For instance, a controller can have read (\textit{R}) and write (\textit{W}) rights only to the submitted reports. Moreover, if the report that has been submitted for approval, it cannot be further modified by the reporter until the end of the approval period. Therefore, during the approval period, the reporter has only read (\textit{R}) right to this particular report. Hence, we should restrict the set of enabled rights depending on a report's state. To address this new behaviour, we refine the corresponding events of the abstract model. Some of the refined $\mathsf{CreateReport}$ and $\mathsf{SubmitReport}$ events are presented in Figure~\ref{fig:devReportronicRef1}.

The model invariants describe the dynamic access policies depending on a report state. These invariants ensure access rights conformity to avoid possible conflicts between the roles. Thereby, it allows us to prove the dynamic data integrity properties within the model.  Some invariants are presented in Figure~\ref{fig:devReportronicRef1}.

\begin{figure}[t!]
	\centering
	\begin{center}
		\begin{small}
\fbox{
	\parbox[center]{15.2cm}{
					\hspace*{0pt} $\textbf{Machine}$ $\mathsf{RMS\_ref1}$ \vspace{1pt}\\
					\hspace*{0pt} $\textbf{Sees}$ $\mathsf{RMS\_c1}$ \vspace{1pt}\\
					\hspace*{0pt} $\textbf{Variables}  \;  \textit{reports},  \textit{report\_state}, \textit{permissions}, ...$ \vspace{1pt}\\
					\hspace*{0pt} $\textbf{Invariants} \; \dots$ \vspace{2pt}\\
					\hspace*{0pt} $\forall{} rp\qdot{} rp \in{} REPORTS \land{} report\_state(rp)=\textit{VOID} \limp{}  (permissions(Reporter\mapsto{}rp)=\{C\} \land{} permissions(Controller\mapsto{}rp)=\emptyset{} \land{} permissions(Administrator\mapsto{}rp)=\emptyset{})$ \vspace{5pt} \\
					\hspace*{0pt} $\forall{} rp\qdot{} rp \in{} reports \land{} report\_state(rp)=\textit{CREATED} \limp{} (permissions(Reporter\mapsto{}rp)=\{R,W,D\} \land{} permissions(Controller\mapsto{}rp)=\emptyset{} \land{} permissions(Administrator\mapsto{}rp)=\emptyset{})$\vspace{5pt} \\
					\hspace*{0pt}  $\forall{} rp\qdot{} rp \in{} reports \land{} report\_state(rp)=\textit{SUBMITTED} \limp{} (permissions(Reporter\mapsto{}rp)=\{R\} \land{} permissions(Controller\mapsto{}rp)=\{R,W\} \land{} permissions(Administrator\mapsto{}rp)=\emptyset{})\vspace*{5pt}$\\
					\hspace*{0pt}  $\forall{} rp\qdot{} rp \in{} reports \land{} report\_state(rp)=\textit{APPROVED} \limp{} (permissions(Reporter\mapsto{}rp)=\{R\} \land{} permissions(Controller\mapsto{}rp)=\{R\} \land{} permissions(Administrator\mapsto{}rp)=\{R,W\})\vspace*{2pt}$ \\
\hspace*{0pt} $ \dots$
\vspace{1pt}\\
					\hspace*{0pt} $\textbf{Events}... \vspace*{2pt}$   \\
						\hspace*{0pt} $\mathsf{CreateReport}$ \textbf{refines} $\mathsf{CreateReport}  \;\widehat{=}$ \vspace{1pt} \\
						\hspace*{5pt} $\textbf{any} \; \; rp$\\
						\hspace*{5pt} $\textbf{where} \; \; ...$\\
						\hspace*{5pt} $C \in permissions(Reporter, rp)$\\
						\hspace*{5pt} $\textbf{then}$\\
						\hspace*{5pt} $\textit{reports} \bcmeq{}  \textit{reports} \bunion{} \{rp\}$\\
						\hspace*{5pt} $report\_state(rp)   \bcmeq{} \textit{CREATED}$\\
						\hspace*{5pt} $permissions(Reporter, rp)   :=  \{R,W,D\}$\\
						\hspace*{5pt} $\textbf{end}\vspace{1pt}$\vspace{2pt} \\
						\hspace*{0pt} $\mathsf{SubmitReport}\;$ \textbf{refines} $\mathsf{SubmitReport}  \;\widehat{=}$\vspace{2pt} \\
						\hspace*{5pt} $\textbf{any} \; \; rp$\\
						\hspace*{5pt} $\textbf{where} \; \; ...$\\
						\hspace*{5pt} $R \in permissions(Reporter, rp)$\\
						\hspace*{5pt} $\textbf{then}$\\
						\hspace*{5pt} $report\_state(rp)   \bcmeq{} \textit{SUBMITTED}$\\
						\hspace*{5pt} $permissions   :=  (permissions \ovl (\{ Reporter \mapsto rp \mapsto \{R\} \} \bunion{}  \{Controller \mapsto rp \mapsto \{R,W\} \})) $\\
						\hspace*{5pt} $\textbf{end}\vspace{1pt}$\\
						...\\
						$\textbf{end}$\\
			}}
		\end{small}
		\caption{The machine $\mathsf{RMS\_ref1}$}
		\label{fig:devReportronicRef1}
	\end{center}
\end{figure}

\paragraph{\textbf{Further Refinements.}}
In the subsequent refinement steps, we augment the specification with further details. In particular, we introduce  for each report a \textit{time window} to model certain periods when a periodic report can be created. We also elaborate on the event $\mathsf{ModifyReport}$ to model the possible changes of a report before its submission. In particular, we define the notion of report timestamps to keep track on time.

Next, we populate our model with the users and introduce a number of inter-relationships between the system users and their roles as well as relationships between the users. For instance, a reporter sends a report for approval to her/his associated controller while the controller sends the approved report to the associated administrator.

As a result of the described refinement chain, we arrive at a final model of RMS. We specify and verify dynamic access control via allowed rights on data according to the system policies.

\paragraph{\textbf{Summary.}}
A development of RMS discussed above follows the certain strategy. The proposed approach can be summarised as follows:
\begin{enumerate}
	\item Define main roles and their functions associated with the system. Represent actors as roles, functions as use cases and create a use case model of the system.
	\item Create an activity diagram representing the intended workflow.
	\item For each use case define the basic access rights required to execute this particular function.
	\item Create a state diagram representing how execution of each function changes the state of the data.
	\item Using the created state diagram, create an abstract specification in Event-B that defines the state of data and the corresponding state transitions.
	\item Using the use case diagram, refine the abstract specification to define roles and the corresponding access rights.
	\item Using the activity diagram, refine the specification to represent the workflow and time-dependant properties of dynamic state-based RBAC.
\end{enumerate}

\section{Conclusions}
\label{sec:concl}
\paragraph{\textbf{Related Work.}}
Significant amount of work has been done on integrating graphical and formal specification techniques to support system development (see e.g., \cite{CunhaCRB11,KimC00,LedangS02,SnookB08}). In particular, the combination UML with Petri Nets is discussed in \cite{CunhaCRB11}. The paper shows a method for translating UML sequence diagrams to Petri nets and verifying deadlock freeness,  reachability, safety and liveness properties. The work \cite{KimC00} presents a framework for integration UML with Object-Z. Various kinds of UML diagrams are used to specify the system from different concerns during the requirements elicitation and analysis stage. Then the captured information is used to develop a complete Object-Z specification. In our work, we follow the same idea and consult the graphical models while creating Event-B specifications.
In \cite{SnookB06,SnookB08} UML-B -- a graphical formal modelling notation  -- has been proposed to support class diagrams and state machines concepts within the Event-B development.
However, the goal of our research is to consider possible combination UML and Event-B modelling in the context of access control model and RBAC policies.

The topic of combining graphical domain-specific notation with formal Event-B models has also been explored in a number of works focusing on modelling dependable systems \cite{ProkhorovaLT15,ProkhorovaT12,SereT99,SereT99FM,Troubitsyna03IPDPS,Troubitsyna08}. The overall goal pursued in these work were to facilitate construction of a formal model by relying on a suitable graphical notation. 

The importance of RBAC visualization has been recognized by Jaeger and Tidswell \cite{JaegerT01}.
A body of research done on applying UML to describe and analyse RBAC policies  \cite{HofrichterGS13,RayLFK04,SunFR11}.  A number of works uses UML and OCL based domain specific language to design and validate  the access control model.
For instance, in the work  \cite{HofrichterGS13} the authors applied UML and OCL to discover and eliminate undesired policy properties, which do not meet the security requirements. In \cite{SunFR11}  UML is also used to describe security properties. In contrast to our work, here the authors transform UML models to Alloy for analysis purpose.

There is a number of works that address the policy analysis and verification issues related to RBAC model. For instance, the problem of inconsistent access control specifications is studied  in \cite{ShafiqMJG05}. To verify the correctness of event-driven RBAC policies a Petri-Net based framework has been applied.
Similar to our work, Event-B has been applied to model and analyse access control policies \cite{AkeelFPGW16,Hoang2008}. Akeel  et al. \cite{AkeelFPGW16} have studied the problem of data leakage threats in the access control model.
They used Event-B and associated refinement approach to formalise the requirements over the specific policy elements that satisfy Confidentiality, Privacy and Trust properties.  In our work we use Event-B to formulate and prove the dynamic data integrity properties.
The challenge of integrating RBAC into systems modelling and verification has been addressed also by Bena{\"{\i}}ssa et al.  in \cite{BenaissaCM07}. They  start from a security policy description given in a Prolog-like formalism (OrBAC), and refine the description into an Event-B model capturing the system-specific activities to be verified under the policy. 

The basic RBAC model has been extended in a variety of ways \cite{AbdunabiARF13,FuchsPS11,RayKY06}.
For instance, the  problem of spatio-temporal RBAC model is discussed in \cite{AbdunabiARF13}. The authors  considered role-based access control policies under time and location constraints. Moreover, they demonstrated how the proposed model can be represented and analysed using UML and OCL. Ray et al. \cite{RayKY06} proposed location-aware RBAC model  that incorporates location constraints in user-role activation.
In our work we consider dynamic, state-dependent constraints within the access control model. In \cite{LPT17} the interactions between agents have been studied using goal-oriented perspective. In this work, the roles were defined as agent capabilities to perform certain tasks.

In our previous work \cite{PereverzevaBFLT14} to facilitate development in Event-B we relied on the Event Refinement Structure approach (ERS). This approach augments Event-B refinement with a graphical notation that allows the designers to explicitly represent the relationships between the events at different abstraction levels as well as define the required event sequences in a model. In current work we used graphical  models as a middle hand to construct Event-B specifications.

Verification of data integrity and data consistency properties in Event-B framework has also been investigated in \cite{PereverzevaLTHP13} in the context of cloud data base. Moreover, in \cite{PereverzevaTL12,PereverzevaTL13} we verified by proofs correctness and safety of consistent updates of patient data records. However, the  current work is mainly focused on possible combination UML and Event-B  to specify and verify dynamic access control.

\paragraph{\textbf{Discussion.}}
In this paper, we have discussed a problem of an integrated modelling of dynamic access rights. Our approach aimed at combining graphical modelling and formal specification in Event-B.  The used graphical models to define in a structured way the roles associated with the system, use cases and workflow. The graphical models were used as a middle hand to construct formal models in \linebreak Event-B.

Formal modelling in Event-B allowed us to rigorously define and verify the dynamic access rights and formulate and prove the dynamic data integrity properties.
In this paper, we merely consulted the graphical models to create the corresponding specifications. Currently, we are working on defining graphical and formal specification patterns for representing the dynamic access rights. Such patterns would allow us to automatically translate graphical models into the corresponding formal models and consequently enable the formal verification of dynamic data integrity properties.

Moreover, as a part of our future work, we are also planing to further extend the current work and consider a more complex setting of access model. In particular, it would be interesting to investigate the situation when the  users can get simultaneous or partial access to some  parts of a data entity depending on their roles and resource states.

\vspace*{-0.2cm}
\bibliographystyle{eptcs}
\vspace*{-0.2cm}
\bibliography{refs}

\end{document}